# Correlated polarization dependences between surface-enhanced resonant Raman scattering and plasmon resonance elastic scattering showing spectral uncorrelation to each other


*Tamitake Itoh[1]\*, Yuko S. Yamamoto[2]*

[1]Nano-Bioanalysis Research Group, Health Research Institute, National Institute of Advanced Industrial Science and Technology (AIST), Takamatsu, Kagawa 761-0395, Japan

[2]School of Materials Science, Japan Advanced Institute of Science and Technology (JAIST), Nomi, Ishikawa 923-1292, Japan




We investigated the origin of the identical polarization angle dependences between surface-enhanced resonant Raman scattering (SERRS) and plasmon resonance for two types of single silver nanoparticle aggregates. The first type (Type I), in which the SERRS spectral envelopes are similar to the plasmon resonance elastic scattering spectra, shows the identical polarization



dependence between the SERRS and plasmon resonance. The second type (Type II), in which the SERRS envelopes largely deviate from the plasmon resonance, "also" exhibits identical polarization dependence. Scanning electron microscopy (SEM) observations indicated that these aggregates were dimers. Thus, this unintuitive result was examined by calculating the electromagnetic (EM) enhancement by changing the morphology of the dimers. The calculation revealed that Type I of dimer generates SERRS directly by superradiant plasmons. The Type II of dimer generates SERRS indirectly by subradiant plasmons, which receive light energy from the superradiant plasmons. This indirect SERRS process clarifies that the interaction between the superradiant and subradiant plasmons results in an identical polarization dependence between SERRS and plasmon resonance for Type II of dimers.

## I. INTRODUCTION

The Raman scattering cross-section of a molecule located inside a nanogap or junction of a plasmonic nanoparticle (NP) aggregate exhibits an enhancement factor of up to $10^{10}$, enabling single-molecule (SM) Raman spectroscopy under resonant conditions.[1-5] These phenomena are called surface-enhanced resonant Raman scattering (SERRS), and their locations are called hotspots (HSs). In addition to the SM sensitivity of SERRS at HSs, HSs have recently received considerable attention because several assumptions in conventional spectroscopies are not applicable within the HSs, resulting in various interesting phenomena, such as strong coupling (breakdown of weak coupling approximation), ultrafast surface-enhanced fluorescence (breakdown of Kasha's rule), vibrational pumping (deviation from thermal equilibrium), and the field gradient effect (breakdown of dipole approximation).[1,6-15] The origin of these phenomena is the extremely small plasmonic mode volume of HSs around several to several tens cubic



nanometers.[1,7,8] Thus, the relationship between these phenomena and plasmon resonance needs to be quantitatively clarified from various viewpoints to understand these phenomena.

The relationship in polarization dependence between SERRS and plasmon resonance has been studied to clarify a SERRS mechanism from the viewpoint of electromagnetic (EM) enhancement using various plasmonic systems.[16-21] A Raman process is composed of an excitation transition and a Raman de-excitation transition. Thus, the EM enhancement factor of SERRS is described as a product of the excitation enhancement factor $F_R(\omega_{ex})$ and Raman scattering enhancement factor $F_R(\omega_{em})$ by plasmon resonance as follows:

$$F_R(\omega_{ex}, \mathbf{r})F_R(\omega_{em}, \mathbf{r}) = \left|\frac{E_{loc}(\omega_{ex}, \mathbf{r})}{E_I(\omega_{ex})}\right|^2 \times \left|\frac{E_{loc}(\omega_{em}, \mathbf{r})}{E_I(\omega_{em})}\right|^2 , (1)$$

where $E_I$ and $E_{loc}$ indicate the amplitudes of the incident and enhanced local electric fields, respectively; $\omega_{ex}$ and $\omega_{em}$ denote the frequencies of the incident and Raman-scattered light, respectively; and $\mathbf{r}$ is the position of a molecule in a HS.[4] Equation (1) indicates that SERRS exhibits the same polarization dependence on plasmon resonance, which generates $F_R(\omega_{ex}, \mathbf{r})F_R(\omega_{em}, \mathbf{r})$ in Equation (1).[16-18] Such plasmon resonance spectra have been confirmed to be similar to the SERRS spectra due to the spectral modulation by $F_R(\omega_{em}, \mathbf{r})$ .[22-24]

In this study, we found that silver NP aggregates which plasmon resonance spectra largely deviate from the SERRS spectra exhibit the same polarization dependence on the plasmon resonances. One might expect that their polarization dependences are uncorrelated because $F_R(\omega)$ does not generated by the observed plasmon resonance. This result might require the modification of EM mechanism; thus, it needs to be clarified. We systematically compared the polarization dependence of SERRS and plasmon resonance for two types of aggregates, which were observed to be dimers by scanning electron microscopy (SEM) images. The first type (Type I) of dimer,



whose SERRS spectral envelopes were similar to the plasmon resonance spectra, showed identical polarization dependence between SERRS and plasmon resonance. The second type of dimer (Type II), in which the SERRS envelopes largely deviated from the plasmon resonance, also exhibited an identical polarization dependence. These results were examined by numerical calculations based on electromagnetism by changing the morphology of the dimers. The calculations indicated that the first type of dimer was symmetric and directly generated EM enhancement by dipole–dipole (DD) coupled (superradiant) plasmons. The second type of dimers is asymmetric and indirectly generates EM enhancement by dipole–quadrupole (DQ) coupled (subradiant) plasmons receiving light energy from DD-coupled plasmons. This indirect SERRS process occurring through subradiant plasmon results in an identical polarization dependence between the SERRS and superradiant plasmon resonance, which spectra largely deviates from the SERRS spectra.

## II. EXPERIMENT

Colloidal silver NPs (mean diameter ~35 nm, $1.10 \times 10^{-10}$ M) were prepared for the SERRS experiment using the method of Lee and Meisel.[25] The colloidal silver NP dispersion was added to the same amount of R6G aqueous solution ($1.28 \times 10^{-8}$ M) with NaCl (5 mM) and left for 30 min for aggregation to obtain SERRS activity. The final concentrations of the R6G solutions ($6.34 \times 10^{-9}$ M) and NP dispersion ($5.5 \times 10^{-11}$ M) were similar to the reported single molecular SERRS condition, as shown by a two-analyte or isotope technique.[26,27] The sample solution (50 μL) was dropped onto a glass slide plate, sandwiched by a glass cover plate to immobilize the SERRS-active colloidal silver NPs on the plate inside the water film and left for 30 min to stably maintain the NPs on the glass slide plate. The sample plate was then placed under an inverted



optical microscope (IX-71; Olympus, Tokyo, Japan). Note that the present dimers were inside a water film.

Figures 1(a1) and 1(b1) illustrate the elastic light scattering and SERRS polarization measurements of the same aggregate on the glass plate surface, respectively. Figures 1(a2) and 1(b2) show the elastic light-scattering and SERRS images, respectively. The elastic scattering light of single silver NP aggregates was detected by illuminating unpolarized white light from a 50-W halogen lamp through a dark-field condenser (numerical aperture (NA) 0.92). When measuring the elastic scattering light, the NA of the objective lens (LCPlanFL 100×, Olympus, Tokyo) was set to 0.6 for dark-field illumination. The SERRS light of the same aggregate was measured by illuminating a unpolarized excitation green laser beam (2.33 eV (532 nm), 3.5 W/cm$^2$, Depolarizer DEQ-2S SigumaKoki) from a CW Nd3+: YAG laser (DPSS 532, Coherent, Tokyo) on the sample plate through another objective lens (5×, NA 0.15, Olympus, Tokyo). When measuring the SERRS light, the NA of the objective lens was increased to 1.3 to efficiently correct the SERRS light. The elastic scattering and SERRS spectra of the single aggregates were measured by selecting one spot on the image plane using a pinhole in front of a polychromator equipped with a thermoelectrically cooled charge-coupled device (CCD) assembly (DV 437-OE-MCI, Andor, Japan).

To measure the dependence of the detection polarization angle ($\theta$) on elastic scattering and SERRS light, both lights were detected using the common polarizer in Figure 1. The SERRS-active silver NPs were always aggregated NPs.[19] If we selected the SERRS-active silver NP aggregates showing dipolar plasmon resonance with maxima of 1.7–2.1 eV and maximum scattering cross-section < 0.1 μm$^2$, they were always dimers.[19] Colloidal gold NPs (mean diameters of 60, 80, and 100 nm; EMGC40, Funakoshi, Japan) were used to convert the scattering intensities to cross-sections. The detailed procedure for conversion is provided elsewhere.[28]



Figure 1

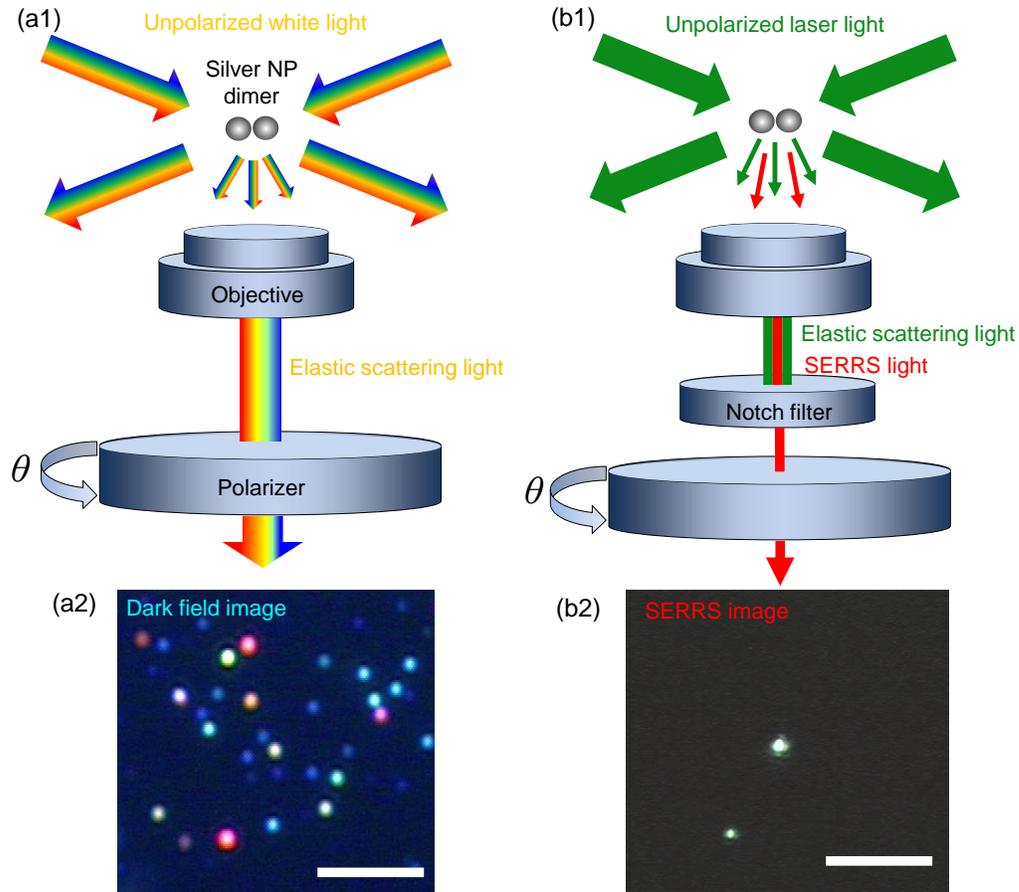

Figure 1. **(a1)** Dark-field illumination of a silver NP dimer. The dimer placed on the cover glass is excited from above, and forward elastic scattered light is detected. The $\theta$ dependence of the scattered light is measured by rotating the polarizer. **(a2)** Dark-field image of NPs and NP aggregates on the cover glass. The single light spot is selected by a pinhole to measure the elastic scattering spectra of a single dimer. Scale bar: 5 μm. **(b1)** SERRS excitation of a NP dimer. The dimer is excited from above, and forward SERRS light is detected. The $\theta$ dependence of the SERRS light is measured by rotating the common polarizer. **(b2)** SERRS image of NP aggregates on the cover glass. The scale bar is 5 μm.

## III. RESULTS AND DISCUSSION



We investigated the relationship between the plasmon resonance maxima in the $\sigma_{sca}(\omega)$ spectra and SERRS spectral envelopes.[29] In this study, we call SERRS with surface enhanced fluorescence, which is broad background emission of SERRS, simply SERRS. The relationship between plasmon resonance maximum energy $\hbar\omega_P$ and SERRS spectral maximum energy $\hbar\omega_S$ for many dimers exhibited two trends in this relationship.[29] One is the correlation between $\hbar\omega_P$ and $\hbar\omega_S$. We call this correlation "Type I". The typical example is indicated in Figure 2(a1). Second, the values of $\hbar\omega_S$ remain around a certain value; even the values of $\hbar\omega_P$ change significantly. We call this lack of correlation "Type II". The typical example is exhibited in Figure 2(c1). There is also an intermediate type between Types I and II as shown in Figure 2(b1).

We compared the $\theta$ dependence of the $\sigma_{sca}(\omega)$ spectra and the SERRS intensities of dimers exhibiting Type I, intermediate type, and Type II by rotating the polarizer, as shown in Figures 1(a)–1(c). Figures 2(a1)–2(c1) show the spectral relationships between $\sigma_{sca}(\omega)$ and SERRS for Type I, intermediate type, and Type II. The blue and black $\sigma_{sca}(\omega)$ spectra were measured using the values of $\theta$ corresponding to the maximum and minimum $\sigma_{sca}(\omega_P)$, respectively. The SERRS intensity reaches a maximum at the $\theta$ corresponding to the maximum $\sigma_{sca}(\omega_P)$. Figures 2(a2)–2(c2) show the contour maps for the $\theta$ dependence of the $\sigma_{sca}(\omega)$ spectra of Figures 2(a1)–2(c1). The plasmon resonance maxima at ~2.0 eV, as indicated by the black open circles, exhibit $\theta$ dependence of the dipole modes, indicating that these resonances are superradiant modes. Spectral dips were observed around 2.4 eV, as indicated by the dashed black open circles, at $\theta$ yielding the maximum $\sigma_{sca}(\omega_P)$. These dips around 2.4 eV become the deepest at the $\theta$ yielding the maximum $\sigma_{sca}(\omega_P)$ at ~2.0 eV. Figures 2(a3)–2(c3) show the $\theta$ dependence of superradiant resonance maxima and that of the averaged SERRS intensities. The $\theta$ dependences of the superradiant resonance maxima are always identical to those of the SERRS intensities for Type I,



intermediate type, and Type II. However, the depolarization ratios of the SERRS were better than those of $\sigma_{\text{sca}}(\omega_{\text{p}})$ for Type II, as shown in Figure 2(c3). These properties are discussed in the calculation section.

Figure 2

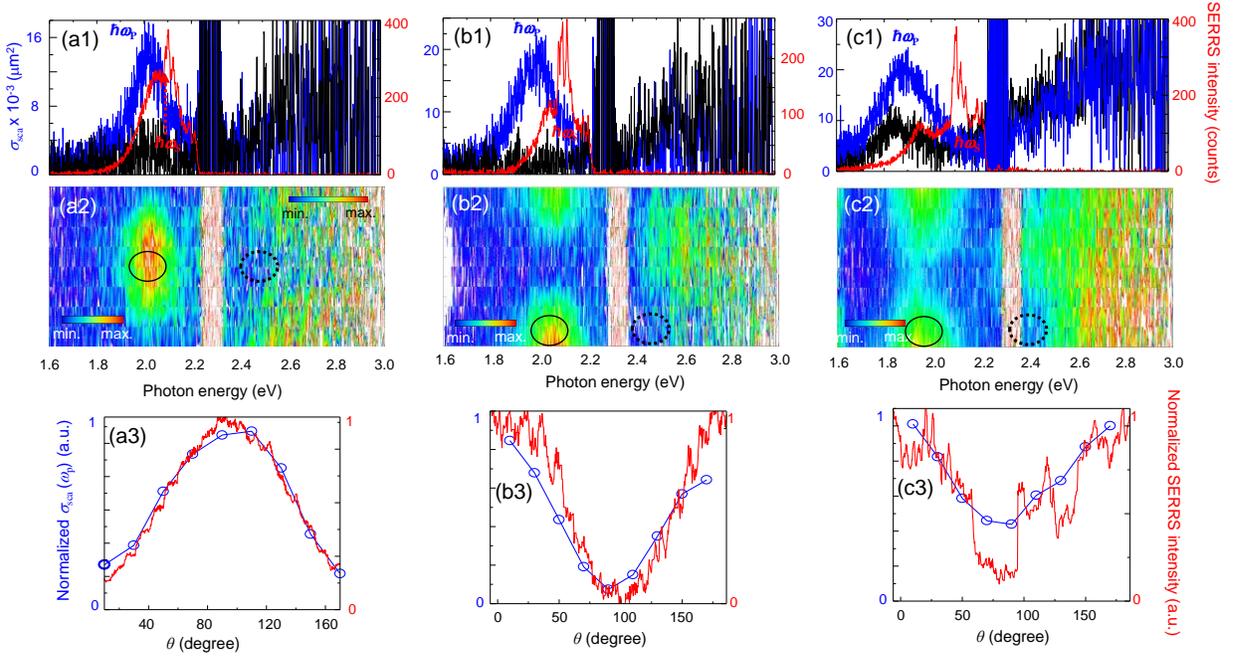

Figure 2. (a1)–(c1) $\sigma_{\text{sca}}(\omega)$ spectra (blue and black curves) and SERRS spectra (red curves) of single dimers exhibiting Type I (a1), intermediate type (b1), and Type II (c1) with unpolarized excitation. The blue and black $\sigma_{\text{sca}}(\omega)$ spectra were measured using the $\theta$ showing the maximum and minimum $\sigma_{\text{sca}}(\omega_{\text{p}})$, respectively. (a2)–(c2) $\theta$ dependence of the $\sigma_{\text{sca}}(\omega)$ spectra of single dimers exhibiting Type I (a2), intermediate type (b2), and Type II (c2) with unpolarized excitation and polarized detection. The spectral maxima and dips are indicated by solid and dashed circles, respectively. (a3)–(c3) $\theta$ dependence of $\sigma_{\text{sca}}(\omega_{\text{p}})$ and SERRS intensity of single dimers exhibiting Type I (a3), intermediate type (b3), and Type II (c3) with unpolarized excitation.

We measured the $\theta$ dependence of the superradiant plasmon resonance maxima and SERRS intensities for 16 dimers to confirm the common $\theta$ dependence for Types I and II. We defined $\theta$ of the superradiant resonance maximum as $\theta_{\text{Pmax}}$ and $\theta$ of the SERRS intensity maximum as $\theta_{\text{Smax}}$. We then evaluated the relationship between $\theta_{\text{Pmax}} - \theta_{\text{Smax}}$ and $\hbar\omega_{\text{P}} - \hbar\omega_{\text{S}}$. The smaller and larger



values of $\hbar\omega_P - \hbar\omega_S$ indicate Type I and II, respectively, as indicated by the arrow in Figure 3(a). Figure 3(a) shows the relationship between $\theta_{Pmax} - \theta_{Smax}$ and $\hbar\omega_P - \hbar\omega_S$. Their common $\theta$ dependences independent of the values of $\hbar\omega_P - \hbar\omega_S$ are clearly observed, indicating the common $\theta$ dependence between SERRS and superradiant plasmon resonance for both Types I and II.

## Figure 3

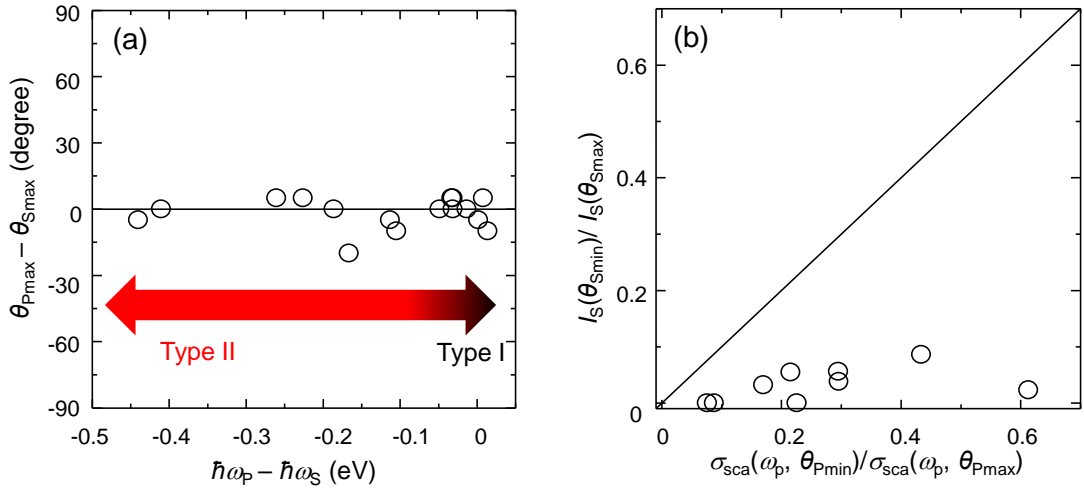

Figure 3. (a) Relationship between the $\theta_{Pmax}$ - $\theta_{Smax}$ and $\hbar\omega_P$ - $\hbar\omega_S$ of single dimers. The degrees of Types I and II are indicated by black and red colored arrows. (b) Relationships in depolarization ratios between $\sigma_{sca}(\omega_p, \theta_{Pmin})/\sigma_{sca}(\omega_p, \theta_{Pmax})$ and $I_S(\theta_{Smin})/I_S(\theta_{Smax})$ of single dimers.

Figures 2(a3)–2(c3) show that the depolarization ratios of $\sigma_{sca}(\omega_p, \theta_{Pmin})/\sigma_{sca}(\omega_p, \theta_{Pmax})$, where $\theta_{Pmin}$ is the $\theta$ for the minimum of $\sigma_{sca}(\omega_p, \theta)$, are always better than $I_S(\theta_{Smin})/I_S(\theta_{Smax})$, where $I_S$ and $\theta_{Smin}$ are the averaged SERRS intensity and $\theta$ for the minimum $\sigma_{sca}(\omega_p, \theta)$, respectively. Figure 3(b) shows the relationship between $\sigma_{sca}(\omega_p, \theta_{Pmin})/\sigma_{sca}(\omega_p, \theta_{Pmax})$ and $I_S(\theta_{Smin})/I_S(\theta_{Smax})$ for the 10 dimers. All data exhibit $\sigma_{sca}(\omega_p, \theta_{Pmin})/\sigma_{sca}(\omega_p, \theta_{Pmax}) >> I_S(\theta_{Smin})/I_S(\theta_{Smax})$, confirming that the depolarization ratios of the SERRS are better than superradiant plasmon resonance.



Here, we discuss the possible mechanism of the identical $\theta$ dependences for Types I and II based on the interaction between superradiant and subradiant plasmons.[30,31] For Type I, the identical $\theta$ dependence between superradiant plasmon resonance and SERRS indicates that the Raman signal is directly enhanced by superradiant plasmon resonance at a HS. Thus, for Type I, the Raman excitation and de-excitation rates were increased by the superradiant resonance. In other words, $F_R(\omega_{ex})$ and $F_R(\omega_{em})$ in Equation (1) are determined by the superradiant resonance. For Type II, the identical $\theta$ dependence between superradiant resonance and SERRS indicates that the SERRS process should be modified. We consider that subradiant resonance may play a major role in the EM enhancement in Equation (1) for Type II. Excitation light energy is first received by superradiant resonance and then transferred to the subradiant resonance with near-field interaction. Then, the rates of Raman excitation and de-excitation are increased by $F_R(\omega_{ex})$ and $F_R(\omega_{em})$ of the subradiant resonance, respectively. Finally, SERRS light is emitted through the superradiant resonance through the inverse process. The receiving and emitting of light by superradiant resonance may be the reason for the identical $\theta$ dependence of superradiant resonance and SERRS for Type II. $F_R(\omega_{ex})$ and $F_R(\omega_{em})$ in Equation (1) are determined mainly by the subradiant resonance, resulting in the spectral deviation of the SERRS envelopes from superradiant resonance. The intermediate type, as shown in Figure 2(b), indicates that both superradiant and subradiant resonance contribute to EM enhancement.

We reported that the morphologies of dimers showing Types I and II are symmetric and asymmetric, respectively, using one-to-one observations between the SERRS spectra, plasmon resonance spectra, and SEM images.[29] Figures 4(a) and 4(b) show typical examples of Types I and II, respectively. The Type I symmetric dimer indicates that the superradiant resonance generating SERRS is a dipole-dipole (DD) coupled resonance. The asymmetry in dimers showing Type II



indicates that subradiant resonance is the coupled plasmon resonance between the dipole of smaller NP and quadrupole of larger NP, namely, dipole-quadrupole (DQ) coupled plasmons. Figures 4(c) and 4(c) illustrate the charge distributions of the DD- and DQ-coupled plasmons, respectively.

## Figure 4

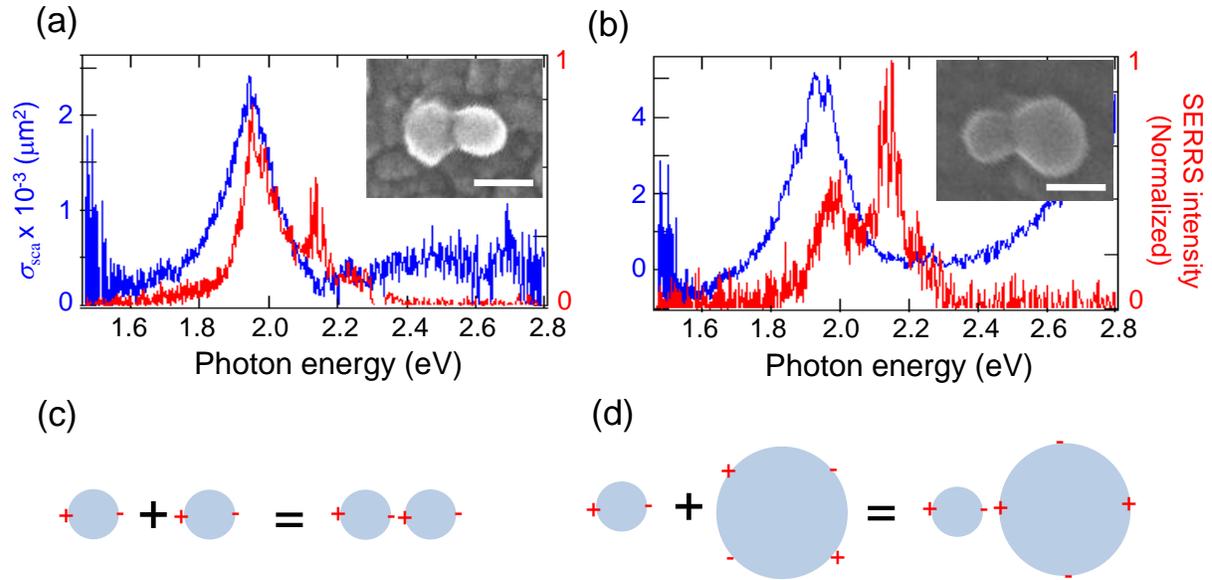

Figure 4. (a) and (b) SERRS spectra (red lines) and $\sigma_{sca}(\omega)$ spectra (blue lines) of dimers, respectively. The inset shows the SEM images of the dimers. Scale bars are 50 nm. (c) and (d) Charge distributions of DD and DQ coupled resonances for symmetric and asymmetric dimers, respectively.

We analyzed the $\theta$ dependences of dimers showing Types I and II by changing the degree of asymmetry of dimers using electromagnetism. The finite-difference time-domain (FDTD) method (EEM-FDM Ver.5.1, EEM Co., Ltd., Japan) was used for the calculation. The complex refractive index of silver for NP dimers was taken from a previous study.[32] The effective refractive index of the surrounding medium was set to 1.39 from the maximum energy of gold NP with a diameter of 80 nm.[28] The validation of the calculation conditions for reproducing the experimental conditions



has been described elsewhere.[19] Note that several studies have reported the contribution of superradiant and subradiant resonance to EM enhancement.[33-36]

Figures 5(a) and 5(b) illustrate the coordinate system of the dimer and the setup for the excitation of a dimer composed of two spherical silver NPs with diameters $D_1$ and $D_2$, respectively, while maintaining a gap of 1 nm. We calculated the $\sigma_{sca}(\omega)$ and $F_R(\omega)$ spectra for the excitation polarization angles $\theta_{ex}$ parallel and perpendicular to the dimer long axis, as shown in Figures 5(c) and 5(d). The $\theta_{ex}$ parallel and perpendicular to the dimer long axis were defined as 0° and 90°, respectively. The phase of $E_{loc}$ at the center gap, as shown in Figures 5(c) and 5(d), was calculated using $\theta_{ex}$ parallel to the dimer long axis because EM enhancement at HSs becomes maximum at this value of $\theta_{ex}$.[19] The initial phase of $E_I$ was set to 180°. The DD couped plasmon resonance, which is assumed to generate Type I, exhibits phase retardation of 90° from the initial phase because DD couped plasmon resonates with the external field. The DQ couped plasmon resonance, which is assumed to generate Type II, exhibits a phase retardation of 180° from the initial phase because DQ couped plasmon resonates with DD couped plasmon. We define the maximum energy in the $F_R(\omega)$ spectrum as $\hbar\omega_F$ as in Figure 5(d), which corresponds to experimentally obtained $\hbar\omega_S$ as in Figure 2(a1).



# Figure 5

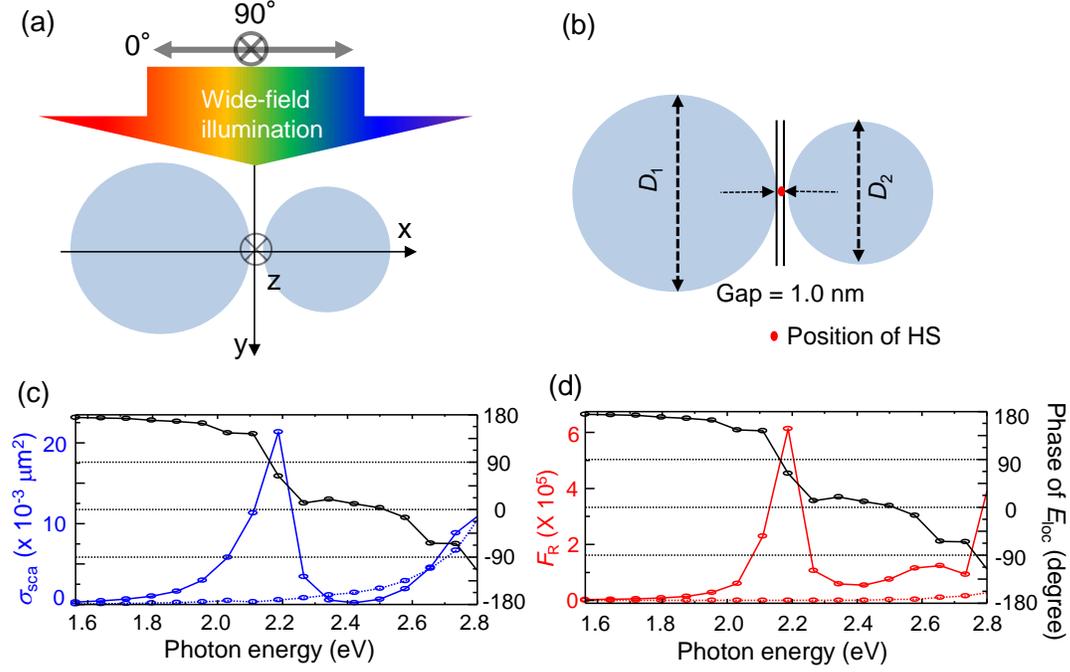

**Figure 5.** **(a)** FDTD calculation setup for electric fields around a single NP dimer by wide-field excitation from the upper side with respect to the coordinate system. Polarization directions of the excitation light are indicated by gray arrows. **(b)** A dimer comprising two NPs with diameters of $D_1$ and $D_2$, respectively. The gap was set to 1 nm. The position of the HS is indicated by the red dot. **(c)** $\sigma_{sca}(\omega)$ spectrum with $\theta_{ex}$ of 0° (blue solid curve) and 90° (blue dotted curve) of a dimer with $D_1$ and $D_2$ of 30 and 50 nm, respectively, with a phase of $E_{loc}$ with $\theta_{ex}$ of 0° (black curve) at the center gap, indicated by the red dot in **(b)**. **(d)** $F_R(\omega)$ spectrum with $\theta_{ex}$ of 0° (red solid curve) and 90° (red dotted curve) of a dimer with $D_1$ and $D_2$ of 30 and 50 nm, respectively, with a phase of $E_{loc}$ with $\theta_{ex}$ of 0° (black curve) at the center gap, indicated by the red dot in **(b)**.

Figures 6(a1)–6(a4) show the $\theta_{ex}$ dependence of $\sigma_{sca}(\omega)$ spectra by increasing $D_2$ from 30 to 120 nm, while keeping $D_1$ fixed at 30 nm. For $\theta_{ex}$ parallel to the dimer long axis, the $\sigma_{sca}(\omega)$ maxima exhibit lower energy shifts from 2.15 to 1.9 eV with increasing $D_1$. These spectral maximum positions are the same as the positions of phase retardation of 90° for $E_{loc}$ against $E_1$, indicating



that these maxima correspond to the DD-coupled resonance. For $\theta_{ex}$ perpendicular to the dimer long axis, $\sigma_{sca}(\omega)$ exhibited spectral broadening with increasing $D_1$. Thus, the depolarization ratios of the elastic scattering $\sigma_{sca}(\omega_p, 90°)/\sigma_{sca}(\omega_p, 0°)$ became worse with increasing $D_1$.



Figure 6

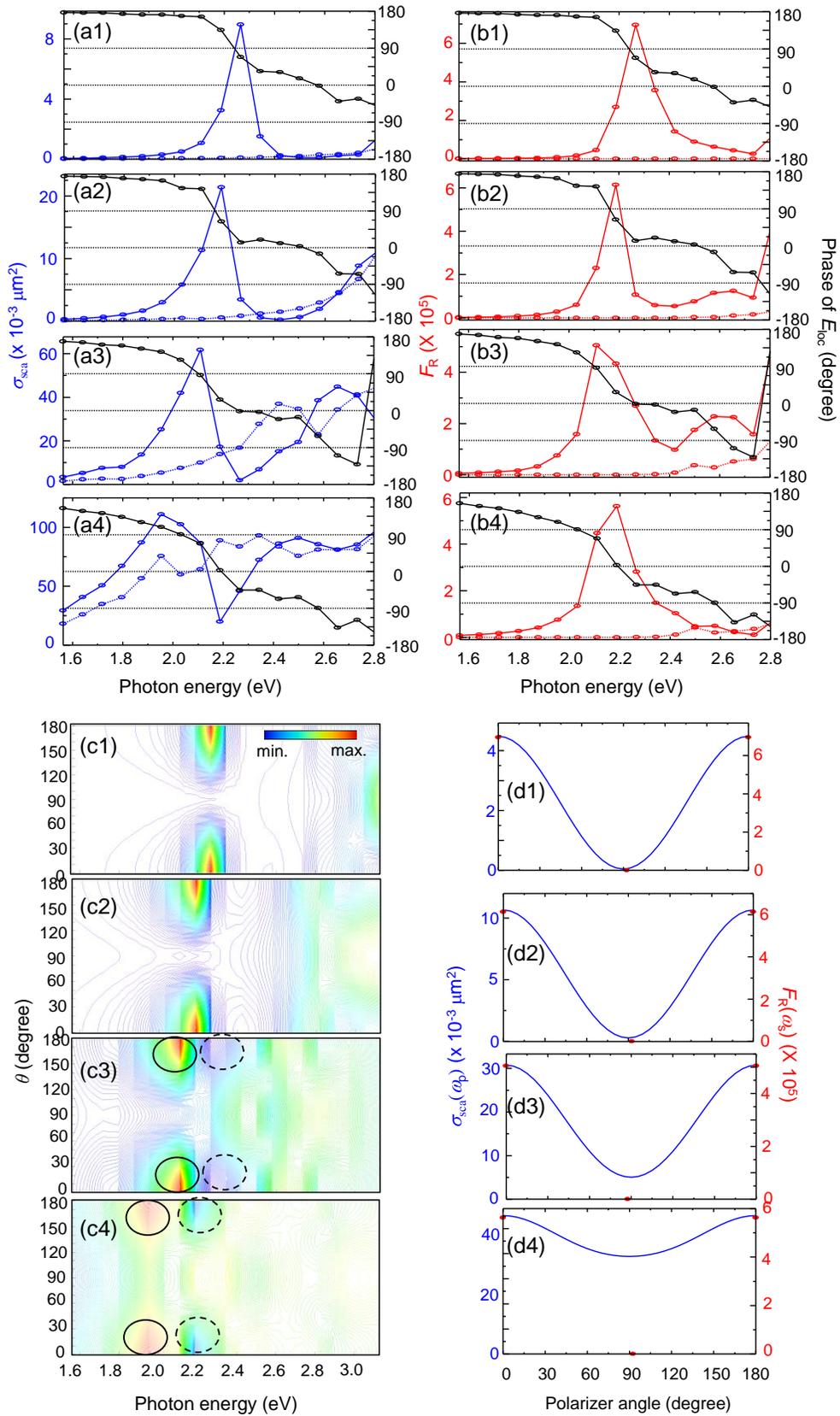



Figure 6. (a1)–(a4) $\sigma_{sca}(\omega)$ spectra with $D_1$ of 30 nm and $D_2$ of 30 (a1), 50 (a2), 80 (a3), and 120 (a4) nm, respectively, with $\theta_{ex}$ of 0° (solid blue curves) with phases of $E_{loc}$ (solid black curves) and $\theta_{ex}$ of 90° (dashed blue curves) to the dimer long axis. (b1)–(b4) $F_R(\omega)$ spectra with $D_1$ values of 30 nm and $D_2$ values of 30 (b1), 50 (b2), 80 (b3), and 120 (b4) nm with $\theta_{ex}$ of 0° (solid red curves) with the phases of $E_{loc}$ (solid black curves) and $\theta_{ex}$ of 90° (dashed red curves) to the dimer long axis. (c1)–10(c4) $\theta$ dependence of $\sigma_{sca}(\omega)$ spectra with $D_1$ of 30 nm and $D_2$ of 30 (c1), 50 (c2), 80 (c3), and 120 (c4) nm, expressed as contour maps. The spectral maxima and dips are indicated by solid and dashed circles, respectively. (d1)–(d4) $\theta$ dependence of $\sigma_{sca}(\omega_p)$ (blue curves) and $\theta_{ex}$ dependence of $F_R(\omega_F)$ (red points) with $D_1$ of 30 nm and $D_2$ of 30 (d1), 50 (d2), 80 (d3), and 120 (d4) nm.

Figures 6(b1)–6(b4) show the $\theta_{ex}$ dependence of $F_R(\omega)$ spectra with increasing $D_2$ while keeping $D_1$ fixed at 30 nm. For $\theta_{ex}$ parallel to the dimer long axis, the $F_R(\omega)$ maxima exhibit lower energy shifts from 2.15 to 2.2 eV with increasing $D_2$. The $F_R(\omega)$ maximum positions for $D_2$ at 30 and 120 nm are the same as the positions of phase retardation at 90° and 180° for $E_{loc}$ against $E_I$, respectively, indicating that these $F_R(\omega)$ maxima correspond to DD and DQ coupled plasmon resonances. This result indicates that $F_R(\omega)$ is generated by the DD and DQ coupled plasmons for symmetric and asymmetric dimers, respectively. Figures 6(a4) and 6(b4) show that the maximum position of $F_R(\omega)$ spectra of $D_2$ at 120 nm at 2.2 eV corresponds to the dip in $\sigma_{sca}(\omega)$ spectrum, indicating that light energy transfer from DD to DQ coupled plasmons generates $F_R(\omega)$.[30,31] For $\theta_{ex}$ perpendicular to the dimer long axis, $F_R(\omega)$ becomes negligible, showing that this DD and DQ coupled resonance cannot generate EM enhancement inside an HS.[19] Thus, the depolarization ratios of SERRS $I_S(90°)/I_S(0)$ do not change with increasing $D_2$. Thus, $I_S(90°)/I_S(0)$ was better than that of $\sigma_{sca}(\omega_p, 90°)/\sigma_{sca}(\omega_p, 0°)$. Note that the common $\theta_{ex}$ dependences of the DD-coupled resonance and $F_R(\omega)$ for both Type I (Figures 6(a1) and 6(b1)) and Type II (Figures 6(a4) and 6(b4)) are also reproduced in the calculations.



The DQ-coupled resonance appears as the deepest dip in the elastic scattering spectra at the same $\theta$ (not $\theta_{ex}$) as the maxima of the DD-coupled resonance, as shown in Figures 2(a2)–2(c2). We examined this property by calculating symmetric and asymmetric dimers. To reproduce Figures 2(a2)–2(c2), the $\theta$ dependences of $\sigma_{sca}(\omega)$ spectra were expressed as contour maps, as shown in Figures 6(c1)–6(c4). These contour maps were created by superposing two contour maps with $\theta_{ex}$ values of 0° and 90° to reproduce the unpolarized excitation and polarized detection conditions. Figures 6(c3) and 6(c4) show that as $D_2$ increases from 30 to 120 nm while keeping $D_1$ fixed at 30 nm, dip structures appear, as indicated by dashed open circles on the higher energy side of the DD-coupled resonance indicated by solid open circles. The dip becomes the deepest at the same $\theta$ as the maxima of the DD-coupled resonance. These properties are consistent with the experimental results in Figures 2(a2)–2(c2), indicating that the light energy transfer from the DD- to DQ-coupled resonance induced the dip structures, and $F_R(\omega)$ reached a maximum at $\theta$ of the maxima of the DD-coupled resonance.

Figures 2(a3)–2(c3) suggest that the depolarization ratios of $F_R(\omega_F)$s are better than those of $\sigma_{sca}(\omega_P)$s. Comparing Figures 6(a1)–6(a4) and Figures 6(b1)–6(b4), it is evident that the depolarization ratios of $F_R(\omega_F)$s are much better than those of $\sigma_{sca}(\omega_p)$. Figures 6(d1)–6(d4) show the $\theta_{ex}$ dependence of $\sigma_{sca}(\omega_p)$ and $F_R(\omega_F)$ with increasing $D_2$ from 30 to 120 nm while keeping $D_1$ fixed at 30 nm. The depolarization ratios of $F_R(\omega_F)$ became better than those of $\sigma_{sca}(\omega_p)$ with increasing $D_2$. The results are consistent with the experiments, indicating that the reason for the better depolarization ratios of $F_R(\omega)$s is as follows. $F_R(\omega)$ is solely generated by DD- or DQ-coupled resonance polarized in the direction parallel to the dimer long axis. By contrast, $\sigma_{sca}(\omega)$ includes spectral components both parallel and perpendicular to the long axis of the dimer. Thus, the depolarization ratios of $\sigma_{sca}(\omega)$s are degraded owing to the inclusion of scattered light from the



dipole resonance polarized perpendicular to the dimer long axis, as shown in Figure 6(a4), supporting the results shown in Figure 3(b).

In the experiments using the 15 dimers shown in Figure 3(a), we evaluated the common $\theta$ dependence for superradiant plasmon resonance and SERRS for both Type I and Type II. We also confirmed this property for the symmetric and asymmetric dimers by changing both $D_1$ and $D_2$ by calculation. Figure 7 shows that all dimers for both Types I and II exhibit common $\theta$ dependencies for $\sigma_{sca}(\omega_P)$ and $F_R(\omega_F)$, which reproduces the experimental results in Figure 3. Thus, we conclude that the two mechanisms of common $\theta$ dependences between SERRS and $\sigma_{sca}(\omega)$ for Types I and II can be explained by the absence and presence of the contribution of the DQ-coupled resonance to SERRS, respectively.

## Figure 7

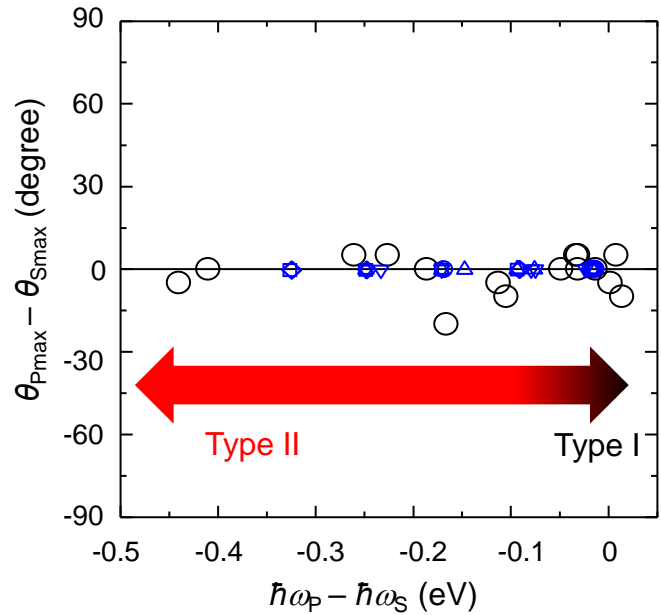

Figure 7. The calculated relationship between $\theta_{Pmax} - \theta_{FMax}$ and $\hbar\omega_P - \hbar\omega_F$ of single dimers, where $\theta_{FMax}$ is the polarization angle that maximizes $F_R(\omega_F)$. The experimental relationship between $\theta_{Pmax}$



$- \theta_{\mathrm{SMax}}$ and $\hbar\omega_{\mathrm{P}} - \hbar\omega_{\mathrm{S}}$ of single dimers. The calculated relationship of single symmetric dimers

($\bigcirc$) and asymmetric dimers with $D_2$ values of 30 ($\square$), 35 ($\triangle$), 40 ($\triangledown$), and 60 ($\diamond$) nm with $\theta_{\mathrm{ex}}$ of 0°.

The diameters of NP symmetric dimers are 30–140 nm. $D_1$ values for asymmetric dimers are 30–

140 ($\square$), 35–140 ($\triangle$), 40–140 ($\triangledown$), and 60–140 ($\diamond$) nm, respectively, with $\theta_{\mathrm{ex}}$ of 0°. The experimental

relationship of single dimers ($\bigcirc$) from Figure 3(a).

We discussed the experimentally obtained $\theta$ dependences of SERRS and elastic scattering based

on DD- and DQ-coupled resonance using $\sigma_{\mathrm{sca}}(\omega)$ and $F_{\mathrm{R}}(\omega)$ spectra. The DD- and DQ-coupled

resonance maxima are attributed to the phase retardations of $E_{\mathrm{loc}}$ against $E_{\mathrm{I}}$ at HSs by FDTD

calculations. To confirm this attribution, we examined the radiation patterns of Type II dimers to

examine the contribution of DQ-coupled resonance to SERRS radiation. Type I dimers have

symmetric structures, and SERRS is generated by DD-coupled resonance. Thus, the radiation

pattern from the HSs exhibits dipole properties. In contrast, Type II dimers have asymmetric

structures, and their SERRS is generated by DQ-coupled resonance. Thus, the radiation pattern

from the HSs may deviate from the dipolar properties, reflecting the destructive interference

between the DD- and DQ-coupled resonances.

Figure 8(a) shows the spectra of $\sigma_{\mathrm{sca}}(\omega)$ and $F_{\mathrm{R}}(\omega)$, and the phase of $E_{\mathrm{loc}}$ at HSs with $\theta_{\mathrm{ex}}$ of 0° for

an asymmetric dimer, which is an intermediate type between Type I and II. The position of $\hbar\omega_{\mathrm{F}}$ is

located between a phase retardation of 90° and 180°. We analysed the radiation patterns for phase

retardations of 90° and 180°, indicated by two vertical black lines in Fig. 8(a), corresponding to

DD and DQ coupled resonances. The charge distributions of the DD-coupled plasmons in Figure

4(c) indicate that the dimer exhibits a dipole radiation pattern. Figure 4(d) illustrates the charge

distribution of DQ-coupled plasmons. We used the generalized Kerker effect on the DD- and DQ-



coupled resonance to analyze the radiation patterns.[37] Figure 8(b) shows the destructive interference of the radiation field between the quadrupole and dipole when the radiation is directed toward the larger NP side. Figure 8(b) also shows that constructive interference occurs when the radiation is directed toward the smaller NP side. Such destructive and constructive interference bends the dipole-like radiation, as shown in Figure 8(b). This radiation pattern of DQ-coupled resonance has been reported for plasmon-enhanced second-harmonic generation.[38-40] We calculated the near electric fields around the dimer to examine the radiation patterns for each phase retardation. Figures 8(c1) and 8(c2) show cross-sectional images (y-planes) of the distributions of $E_{loc}$ for each radiation phase. These $E_{loc}$ distributions appear complex, but $E_{loc}$ in Figure 8(c2) shows bending in the distribution in the direction of the smaller NP, as indicated by the green arrows. This property is consistent with the radiation pattern of the DQ-coupled resonance shown in Figure 8(b). We also examined the far-field radiation patterns at each spectral position. Figures 8(d1) and 8(d2) show the polar diagrams of the radiation patterns for the DD- and DQ-coupled resonances. The radiation pattern shown in Figures 8(d1) exhibits dipole features. However, Figure 8(d2) shows bending to the side of the smaller NP, which is evidence of the contribution of DQ-coupled resonance to the radiation pattern.



# Figure 8

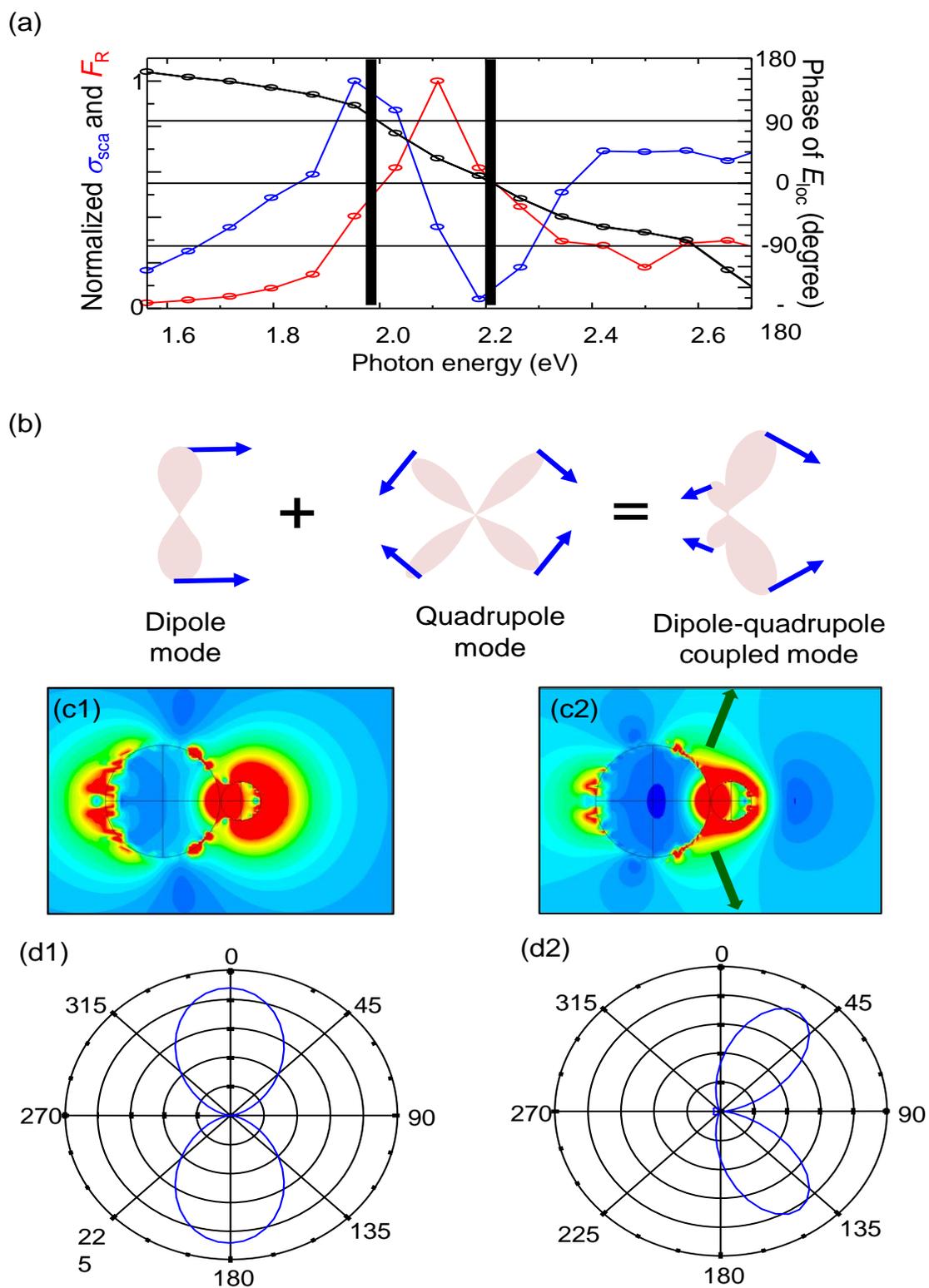

(a)

(b)

Dipole mode + Quadrupole mode = Dipole-quadrupole coupled mode

(c1)   (c2)

(d1)   (d2)



Figure 8. (a) Normalized $\sigma_{sca}(\omega)$ and $F_R(\omega)$ spectra with phases of $E_{loc}$ at the HS of a dimer with $D_1$ of 35 nm and $D_2$ of 100 nm with $\theta_{ex}$ of 0°. The initial phase of $E_1$ is set 180°. The two vertical solid lines indicate positions for phase retardation of 90° (DD-coupled resonance) and 180° (DQ-coupled resonance). (b) Polar diagrams of the radiation patterns of the z plane for dipole, quadrupole, and DQ-coupled resonance. The blue arrows indicate the directions of the electric fields. (c1) and (c2) Distributions of the $E_{loc}$ of the Y-plane of a dimer for DD- and DQ-coupled plasmons at the two spectral positions in (a). (d1) and (d2) Polar diagram of the radiation patterns at the two spectral positions in (a).

Figures 6(b1)–6(b3) show that $F_R(\omega)$ is dominated by DQ-coupled plasmons when the degree of asymmetry in the dimers is increased. The degree of asymmetry also affects the far-field radiation pattern as it deviates from the dipolar radiation pattern. The contribution of the DQ-coupled resonance to the radiation pattern can be evaluated by polar diagrams as the bending of the radiation direction, as shown in Figure 8(d2). Thus, we examined the bending angles of the radiation patterns by changing the degree of asymmetry, defined as $D_1/D_2$. Figure 9 shows the relationship between $D_1/D_2$ and the bending angle $\phi$, as defined in the inset of Figure 9. In symmetric dimers, the values of $\phi$ are always 0°, indicating a dipole radiation. By increasing $D_1/D_2$, $\phi$ increased and saturated at ~35°. This bending, which can be experimentally observed using a Fourier imaging method,[41,42] is useful for evaluating the contribution of the DQ-coupled resonance to SERRS.



# Figure 9

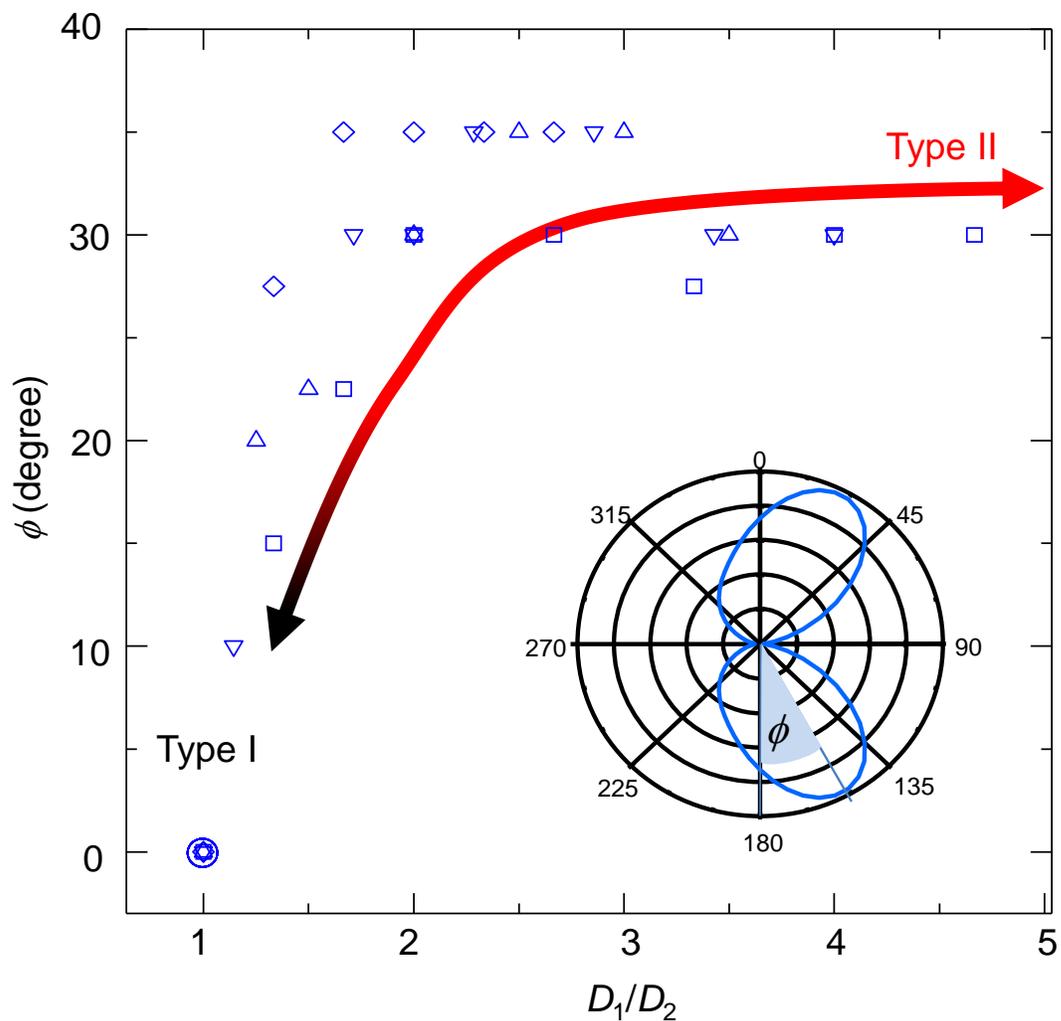

**Figure 9.** Relationships between the $D_1/D_2$ and $\phi$ of single symmetric dimers ($\circ$) and asymmetric dimers with $D_2$ values of 30 ($\square$), 35 ($\triangle$), 40 ($\triangledown$), and 60 ($\diamond$) nm. The diameters of the NP symmetric dimers are 30–140 nm. The $D_2$ values for the asymmetric dimers are 30–140 ($\square$), 35–140 ($\triangle$), 40–140 ($\triangledown$), and 60–140 ($\diamond$) nm, respectively. The arrow indicates the trend of Type I or II dimers. The inset is a definition of $\phi$ of the radiation patterns in the polar diagram.

## IV. SUMMARY



We investigated the identical polarization dependences between superradiant plasmon resonance and SERRS, which the spectral envelopes largely deviated from the superradiant resonance, using single silver NP aggregates. The SEM observations revealed that the aggregates were dimers. Thus, the FDTD calculation of the plasmon resonance and $F_R(\omega)$ was performed by changing the degree of morphological asymmetry in the dimers. The calculations accurately reproduced identical polarization dependences. The analysis of the phase retardations of $E_{loc}$ at HSs revealed that the $F_R(\omega)$ of symmetric and asymmetric dimers is increased by both DD- and DQ-coupled resonances, respectively. The DQ-coupled resonance, which is subradiant, receives excitation light and emits SERRS light through near-field interaction with the DD-coupled resonance. This indirect SERRS process causes identical polarization dependencies between superradiant plasmon resonance and SERRS, whose spectral envelopes largely deviate from superradiant resonance spectra. The contribution of the DQ-coupled resonance to $F_R(\omega)$ was also theoretically evaluated using the deviation of the radiation pattern from that of the DD-coupled resonance. This study indicates the importance of the subradiant resonance of the EM enhancement of various plasmonic HSs composed of NP or nanowire dimers, NPs on mirrors, and NP clusters.[19,43-48]

## AUTHOR INFORMATION


### Corresponding Author

*Corresponding author: tamitake-itou@aist.go.jp


### Author Contributions

The manuscript was written through contributions of all authors. All authors have given approval to the final version of the manuscript.


### Funding Sources





This work was supported by a JSPS KAKENHI Grant-in-Aid for Scientific Research (C) (number 21K04935).